\parskip=10pt

\def\A{{\cal A}}
\def\B{{\cal B}}
\def\G{{\cal G}}
\def\R{{\bf R}}
\def\I{{\cal I}}
\def \d{{\rm d}}
\def\pd{\partial}
\def\sp{\sum\nolimits'}

\def \a{\alpha}
\def \b{\beta}
\def \om{\omega}
\def \ka{\kappa}
\def \z{\zeta}
\def \l{\lambda}
\def \phi{\varphi}

\def \tr {transformation}
\def \com {constant of motion}
\def \coms {constants of motion}
\def \sy {symmetry}
\def \sys {symmetries}
\def \an {analytic}
\def \co {convergen}

\def \qq {\qquad}
\def \en {\eqno}
\def \cd {\cdot}
\def \grad {\nabla}
\def \Ker {{\rm Ker}}
\def \sse {\subseteq}

\def \pn{\par\noindent}
\def \bs{\bigskip}
\def \Ref{\bs\bs\pn{\bf References}\parskip=7 pt\parindent=0 pt}
\def \section#1{\bs\bs \pn {\bf #1} \bigskip}

\def\~#1{\widetilde #1}
\def\.#1{\dot #1}
\def\^#1{\widehat #1}

\baselineskip 0.6 cm
{\nopagenumbers
~ \vskip 4 truecm
{\bf
\centerline {
Convergent Normal Forms of Symmetric Dynamical Systems}
\vskip 2 truecm
\centerline {G. Cicogna}}
\centerline{Dipartimento di Fisica, Universit\`a di Pisa, 
P.za Torricelli 2, I-56126, Pisa, Italy}
{\tt \centerline{E-mail : cicogna@ipifidpt.difi.unipi.it}}
\vskip 4 truecm
\pn
{\bf Abstract.} 
\pn
It is shown that the presence of Lie-point-symmetries of 
(non-Hamiltonian) dynamical systems can ensure the 
convergence of the coordinate transformations which take the dynamical 
sytem (or vector field) into Poincar\'e-Dulac normal form.

\vfill\eject }

~ \vskip 1.5 truecm
\baselineskip 0.53 cm

\section{1. Introduction}

A well known and interesting procedure, going back to the classical work of
Poincar\'e, for investigating \an\ vector fields (VF) $X_f$ 
$$X_f\equiv \sum_{i=1}^n f_i(u){\pd\over{\pd u_i}}\equiv f\cd \grad \qq 
(u\in \R^n)\en(1)$$ 
or the associated dynamical sytems (DS)
$${\d u\over{\d t}}=f(u)\qq u=u(t)\en(2)$$
in a neighbourhood of a stationary point, 
is that of introducing some new coordinates in which the given VF takes 
its ``simplest'' form, i.e. the normal form (NF) (in the sense of 
Poincar\'e-Dulac [1-4]). These coordinate \tr s are usually performed by 
means of recursive techniques: in general the normalizing \tr s (NT) are 
actually purely formal \tr s, and only very special conditions can 
ensure their \co ce and the (local) analyticity of the NF [1-4]. In the 
investigation of these problems, a 
relevant role can be played  by the presence of some \sy\ property 
[5,6] (see also [7]) of the VF $X_f$, i.e. by the presence of some VF $X_g$ 
$$X_g\equiv \sum_{i=1}^n g_i(u){\pd\over{\pd u_i}}\equiv g\cd \grad 
\en(3)$$ 
such that
$$ [X_f,X_g]=0\en(4)$$
In terms of the DS (2), the \sy\ VF $X_g$ provides the Lie generator of a 
(possibly non-linear) Lie-point-symmetry of the DS, and this can be 
conveniently expressed in the form of the Lie-Poisson bracket
$$\{f,g\}_i\equiv (f\cd\grad)g_i-(g\cd\grad) f_i \ =\ 0 \qq (i=1,\ldots,n)
\en(5)$$
In this context, Bruno and Walcher [8] showed that the existence of an \an\ 
\sy\ for a 2-dimensional DS is enough to ensure \co ce of the NT; in [9] the 
\co ce is obtained also for DS of dimension $n>2$, combining the 
existence of \sys\ with other conditions involving also 
the \coms\ of the DS; in [10,11] the role of \sys\ is investigated 
in view of the problem of linearizing the DS.
In this paper, we will discuss some generalizations and some 
results along the same lines.
\vfill\eject

\section {2. Preliminary results}

We will freely use $f$ both to denote the VF $X_f$ (1) and to refer to the DS 
(2); let us introduce the notation
$$f(u)=Au+F(u)\en(6)$$
where $f$ is assumed to be \an\ in a neighbourhood of the stationary point
$u_0=0$, and its linear part $A=(\grad f)(0)$ a 
semisimple (and not zero) matrix. The NF of $f$ will be written
$$\^f(u)=Au+\^F(u)\en(7)$$
(the notation \ $\^\cd$\ will be always reserved to NF; there is no 
danger of confusion if $u$ is used to denote also the ``new'' coordinates), 
and $\^F(u)$ contains the ``resonant terms'' with respect to $A$, i.e. 
the terms such that
$$\^F(u)\in\Ker(\A)\en(8)$$
where $\A$ is the ``homological operator'' defined by
$$\A(h)=\{Au,h\}\en(9)$$
This fact can be conveniently stated in the form:
\pn
{\bf Proposition 1.} Every NF $\^f$ admits the linear \sy\ $g_A\equiv Au$. 
\pn
Let us recall some well known and useful facts. 
\pn
{\bf Lemma 1.} Every \sy
$$g(u)=Bu+G(u)\en(10)$$
of a NF $\^f$ is also a \sy\ of the linear part $Au$ of $\^f$. The analogous 
result is true for the \coms\ of the DS (6): i.e., if a scalar function 
$\mu=\mu(u)$ is such that $\^f\cd\grad\mu=0$ then also $Au\cd\grad\mu=0$. 
\pn
Denoting by $\G_f$ and $\I_f$ the set of the \sys\ and respectively of 
the \coms\ of $f$, we can then write
$$\G_{\^f}\sse\G_{Au} \qq{\rm and} \qq \I_{\^f}\sse\I_{Au}\en(11)$$
This is true not only for \an\ quantities, but also for quantities 
expressed by means of formal power series.
\pn
{\bf Lemma 2.} Given the matrix $A$, the most general NF has the form
$$\^F(u)=\sum_j \mu_j(u) M_ju \qq{\rm with}\qq \mu_j(u)\in\I_{Au}
\qq {\rm and}\qq [M_j,A]=0 \en(12)$$
where the sum is extended to a set of linearly independent matrices 
$M_j$ (the set of these matrices clearly includes $A$). 
\pn
{\bf Lemma 3.} If $f$ admits a linear \sy\ $g_B=Bu$, then the NF $\^f$ 
also admits this \sy . If $f$ admits a (possibly formal)  \sy\ 
$g=Bu+G(u)$ and $B$ is semisimple, then $Bu$ is a \sy\ of the NF $\^f$, or 
-- in other words -- $\^F$ is a NF also with respect to $B$, i.e. 
$\^F\in\Ker(\A)\cap\Ker(\B)$.

The proofs of these Lemmas are well known and can be found, e.g., in 
[7,12-14]. Let us also recall the basic conditions, found by Bruno [1,2], 
and called Condition $\om$ and Condition A, which ensure 
the \co ce of the NT of a given DS. Denoting by $\l_1,\ldots,\l_n$ the 
eigenvalues of the matrix $A$, then the first condition is
\pn
{\sl Condition $\om$}: let $\om_k=\min|(q,\l)|$ for all positive 
integers $q_i$ such that $\sum_{i=1}^n q_i<2^k$ and 
$(q,\l)=\sum_iq_i\l_i\ne 0$: then 
$$\sum_{k=1}^\infty 2^{-k}\ln \om_k^{-1}<\infty$$
\pn
This is a very weak condition, and we explicitly 
assume from now on that it is always satisfied. The other one, instead, 
is a quite strong restriction on the form of the NF. To state this 
condition in its simplest form, let us assume for a moment that there is 
a straight line through the origin in the complex plane which contains 
all the eigenvalues $\l_i$ of $A$, and that there are eigenvalues lying 
on both sides of this line with respect to the origin. Then the condition 
reads
\pn
{\sl Condition A}: there is a coordinate \tr\ changing $f$ to $\^f$, 
where $\^f$ has the form
$$\^f=Au+\a(u)Au$$
and $\a(u)$ is some scalar-valued power series \big(with $\a(0)=0$\big).
\pn
In the case there is no line in the complex plane which satisfies the 
above property, then Condition A should be modified [1] (or even 
weakened: for instance, if there is a straight line through the origin 
such that all the $\l_i$ lie on the same side of this line, then the 
eigenvalues belong to a Poincar\'e domain [1,3] and the \co ce is 
guaranteed without any other condition); but in all the applications 
below, where in particular only linear NF will be ultimately 
concerned, the above formulation of Condition A is enough to cover 
all the cases to be considered, and we can say [1,2] that 
there is a \co t NT if the above conditions are satisfied. Clearly, here 
and in the following, ``\co ce'' stands for ``\co ce in some open 
neighbourhood of $u_0=0$''.

\section{3. Symmetries and \co ce of the NT: a general result}

Let us finally state the first result of this paper. It can be noted that 
quite strong assumptions are needed; but it is known, on the other hand, that 
the \co ce of the NT is quite ``exceptional''. The examples given below 
will show how, thanks to additional \sy\ properties, these assumptions 
can happen to be verified. Let us remark that obviously, for any constant 
$c$, then $cf$ is a (trivial) \sy\ of $f$;  therefore, it is understood 
that when we assume the existence of some \sy\ of $f$ we will always 
refer to {\it nontrivial} \sys , i.e. to \sys\ $g\not= cf$.
\pn
{\bf Theorem 1.} Given the \an\ VF $f$, let us write its NF, according to 
Lemma 2, in the form
$$\^f=Au+\a(u) Au+\sp\mu_j(u)M_ju\equiv Au+\a(u) Au+\^F_1(u)\en(13)$$
where (here and in the following) $\sum'$ is the sum extended to the 
matrices $M_j\not=A$. Assume $\^F_1(u)\not=0$ (otherwise Condition A is 
sufficient of ensure \co ce of the NT), and:
\parskip 0 pt\pn
{\sl a)} assume that $f$ admits an \an\ \sy\ 
$$g=Bu+G(u)\qq {\rm such\ that}\qq B=aA\en(14)$$ 
where $a$ is a (possibly vanishing) constant;
\pn
{\sl b)} assume that the equation
$$\{\^F_1,S\}=0\en(15)$$
for the unknown
$$S=S(u)=\sp\nu_j(u)M_ju \qq {\rm with} \qq \nu_j(u)\in\I_{Au}\qq {\rm and}
\qq \nu_j(0)=0\en(15')$$
has only the trivial solution
$$S=c\^F_1(u) \qq\qq (c={\rm constant}) $$
Then $f$ can be put into NF by means of a \co t NT.
\parskip 10 pt\pn
{\bf Proof}. First of all, if $a=0$ in assumption {\sl a)}, one 
can consider, instead of 
$g$, the \sy\ $g'=f+g$ having linear part $Au$; it is then not 
restrictive to assume $a=1$, i.e. $B=A$. Once $f$ is put into NF $\^f$, 
the \sy\ $g$ will become a (possibly formal) \sy\ $\~g$
$$\~g=Au+\b(u)Au+\sp\nu_j(u)M_ju\qq{\rm with}\qq\b(u),\nu_j(u)\in\I_{Au}
\en(16)$$
this is indeed the most general \sy\ of a NF, thanks to Lemma 1. The \sy\ 
condition $\{f,g\}=0$ in the new coordinates reads $\{\^f,\~g\}=0$;  
evaluating term by term this bracket, one is left with
$$\Big\{\a Au\ ,\ \sp_j \nu_j M_ju\Big\}+\Big\{\sp_j 
\mu_j M_ju\ ,\ \b Au\Big\}
+
\Big\{\sp_j\mu_j M_ju\ ,\ \sp_k\nu_k M_ku\Big\}\ =\ 0$$
or
$$\Big(\sp_j\mu_j(u)M_ju\cd\grad\b-\sp_j
\nu_j(u)M_ju\cd\grad\a\Big)Au\ +\ 
\Big\{\sp_j\mu_j(u)M_ju\ ,\ \sp_k\nu_k(u)M_ku\Big\}\ =\ 0 \en(17)$$
All other terms in fact vanish thanks to Proposition 1 and Lemmas 2 and 
3. Now, in eq.(17), the bracket $\{\ \cd\ ,\ \cd\ \}$ produces, through 
the matrix commutators $[M_j,M_k]$, only terms 
proportional to $M_ju$ (and not to $Au$: this can be easily seen in a 
basis in which $A$ is diagonal), therefore the terms appearing into 
the $\big(\quad\big)$ and the bracket $\{\ \cd\ ,\ \cd\ \}$ are both zero. 
The last bracket has just the form 
$\{\^F_1,S\}$, and therefore assumption {\sl b)} gives $S=c\^F_1$, i.e. 
$\nu_j(u)=c\mu_j(u)$. From the 
vanishing of the first $\big(\quad\big)$ in (17), and using again 
assumption {\sl b)}, one obtains similarly $\b(u)=c\a(u)$, 
and then either $\~g=c\^f$, which is impossible because $g\not=cf$, or 
$$\~g(u)\ =\ Au \en(18)$$
This means that the \tr\ which puts $f(u)$ into $\^f(u)$ transforms 
$g(u)=Au+G(u)$ into $\~g(u)=Au$, therefore the \sy\ $g(u)$ satisfies 
Condition A and there is \co t \tr\ which puts $g(u)$ into NF. Under this 
\co t \tr\ $f(u)$ is transformed into NF $\^f$, as a consequence of the 
last part of Lemma 3.

\pn
{\bf Remark 1.} One can see that the assumption {\sl b)} of Theorem 1 is 
equivalent to the assumption that the NF $\^f$ admits only linear \sys\ 
$Lu$. For a practical point of view (see the Examples below), it is quite 
simpler to verify the property {\sl b)}.

\pn
{\bf Remark 2.} In the particular case that $\^F_1(u)$ has the form
$$\^F_1(u)=\mu(u)Mu\en(19)$$
(with $M\not= A$), then assumption {\sl b)} is actually equivalent to the 
very simple following one: there are no common (\an , formal or fractional) 
\coms\ of the two linear problems
$$\.u=Au\qq{\rm and}\qq\.u=Mu\en(20)$$
Indeed, assume there is some $\ka=\ka(u)\in\I_{Au}\cap\I_{Mu}$, then 
$S=\ka(u)Au\not= c\^F_1$ would satisfy $\{\^F_1,S\}=0$; notice, 
incidentally, that one would also get in this case $\ka(u)\in\I_{\^f}$\ . 
The converse is easily obtained by explicit calculations. This case has 
been already considered in [9]; the result for 2-dimensional 
DS in [8] can be viewed as a particular case of this (see [9] for 
details).

\bs
The apparent difficulty in the application to concrete cases of the 
above results is that, in general, one does not known -- a priori -- the 
NF, and then it seems to be impossible to check if the assumptions of 
Theorem 1 (or even Condition A) are verified by the NF. However, as the 
foregoing Examples will show, other \sy\ properties of the VF may provide, 
once again, the decisive help on this point.

\pn
{\bf Example 1.} Consider a 3-dimensional \an\ DS
$$\.u=f(u)=Au+F(u) \qq {\rm with}\qq A={\rm diag}(1,1,-2)\en(21)$$
with $u\equiv(x,y,z)\in\R^3$ and assume that $f(u)$ possesses the linear 
$SO_2$ \sy\ generated by $Lu\cd\grad$ where
$$L=\pmatrix{0 &1 & 0\cr
             -1 &0 &0 \cr
             0 &0 &0 } \en(22)$$
i.e. $f$ is ``equivariant'' under rotations in the plane $(x,y)$. Putting 
$r^2=x^2+y^2$, this implies that $F(u)$ must be written in the form
$$ F(u)=\phi_0(r^2,z)Au+\phi_1(r^2,z)Iu+\phi_2(r^2,z)Lu \en(23)$$
where $I$ is the identity matrix in $\R^3$.
If we now choose, for instance,
$$\phi_0=0,\qq \phi_1=a_1r^2z+a_2z^3,\qq \phi_2=b\phi_1\en(24)$$
where $a_1,a_2,b$ are constants $\not=0$, then the DS admits also the 
non-linear \sy\
$$G(u)=r^2z(I+bL)u\en(25)$$
Notice that the assumption $a_2\not=0$ ensures that this DS is {\it not} 
a NF, and that the above \sy\ is not trivial.
Then assumption {\sl a)} of Theorem 1 is satisfied. Now, the NF of the 
above DS (21-24) must be of the form
$$\^f=Au+\a(r^2z)Au+\mu_1(r^2z)Iu+\mu_2(r^2z)Lu \en(26)$$
where $\a,\mu_1,\mu_2$ depend only on $\ka=r^2z$, as a consequence of Lemma 3 
(i.e., the equivariance under $SO_2$ is preserved), and of Lemma 2 
(the resonance condition).
We have to look for the solutions $S$ of the equation 
$\{\^F_1,S\}=0$, where the  unknown  $S$ can be written
$$S=\nu_1(u)Iu+\nu_2(u)Lu\qq{\rm with}\qq\nu_1(u),\nu_2(u)\in\I_{Au}\en(27)$$
\big(it is easy to see that no other matrices can appear in the r.h.s. of 
(26-27)\big), and where $\nu_1,\nu_2$ must be functions only of the two 
functionally independent quantities 
$x^2z,xyz\in\I_{Au}$. The condition $\{\^F_1,S\}=0$ gives 
the first-order system of linear partial differential equations
$$\eqalign{(\mu_1 u+\mu_2Lu)\cd\grad\nu_1=\nu_1 u\cd\grad\mu_1\cr
  (\mu_1 u+\mu_2Lu)\cd\grad\nu_2=\nu_1 u\cd\grad\mu_2 }\en(28)$$
Observing that 
$Lu\cd\grad\mu_1=0$ and $Au\cd\grad\mu_1=0=u\cd\grad\mu_1-3z\pd_z\mu_1$ 
and the same for $\nu_1$, i.e. $u\cd\grad\nu_1=3z\pd_z\nu_1$, one gets 
from the first of the (28) the characteristic equation 
for the unknown $\z=\z(u)$ defined by $\nu_1(u)=\z(u)\mu_1(u)$
$${\d x\over y}\ =\ -\ {\d y\over x}\ =\ {\mu_2\over\mu_1}\ 
{\d z\over 3z} \en(29)$$
which shows that $\z$ must be a function of $r^2$ and of some other 
variable of the form $v={\rm arctg}(y/x)+Z(z)$. On the other hand, 
$\mu_1$ is a function of $\ka=r^2z$, and $\nu_1$ is a function 
of the quantities $x^2z,xyz\in\I_A$ 
(possibly also of $x/y$, of course, the only requirement is that $\mu_1,\ 
\nu_1$ are power series in $x,y,z$); it is then easy to see that 
$\z={\rm const}$, i.e. that $\nu_1=c\mu_1$. Proceeding in the same way 
for the other equation in (28), one can conclude that
$$S\ =\ c\^F_1$$
and then also the assumption {\sl b)} in Theorem 1 is satisfied, and 
therefore there is a \co t NT. 

Notice that it was essential in the calculations for the example above
that $\mu_1$ and $\mu_2$ are both $\not=0$, and this is in fact 
guaranteed by the normalizing procedure: indeed, at the lowest order, 
the resonant terms $r^2zIu$ and $r^2zLu$ are orthogonal to $z^3Iu$ 
and $z^3Lu$ (with respect to standard scalar product [3,12,14] introduced 
in the vector space of homogeneous polynomials, where the homological
operator $\A$ is defined), then at the first step 
of the normalization procedure the resonant terms are not changed, i.e. 
one has $\mu_1=a_1r^2z+\ldots$ and $\mu_2=a_1b\ r^2z+\ldots$;
and then -- at any further step of the iteration -- the lower order 
terms are not altered.

\section{4. Symmetries and convergence of the NT: a special case}
 
Let us now come back to the special case that 
$\^F_1(u)$ can be written in the form 
$$\^F_1(u)=\mu(u) Mu\en(30)$$
In Remark 2 we have seen that assumption {\sl b)} of 
Theorem 1 can be replaced by the requirement that there are no 
simultaneous \coms\ of $Au$ and $Mu$. Assume now that, as in Example 1, 
the DS admits not only the non-linear \sy\ $g(u)$ \big(assumption {\sl a)} of 
Theorem 1\big), but also a \sy\ generated by some linear VF $Lu\cd\grad$, and 
that $g(u)$ satisfies
$$\{g,Lu\}=0\en(31)$$
In the NT the \sy\ is conserved step by step [14], therefore 
$g(u)$ also will be 
transformed, once $f$ is in NF $\^f$, into some $\~g(u)$ which is 
symmetric under $Lu$: $\{\~g,Lu\}=0$. From this remark we see that 
it is sufficient to look for the common \coms\ of $Au$ and $Mu$ 
{\it only in the set} of those 
scalar functions $\ka=\ka(u)$ which are left invariant by $Lu$; in other 
words, we can conclude that if the set of these simultaneously invariant 
functions contains only trivially constant numbers, i.e. 
$$\I_{Au}\cap\I_{Mu}\cap\I_{Lu}=\R \en(32)$$
then no other non-linear \sys\ are allowed, and $\~g(u)$ becomes necessarily 
$\~g(u)=Au$; then, using similar arguments as above, the \co ce of the NT is 
guaranteed. In conclusion, we can state: 
\pn 
{\bf Theorem 2.} Assume that $f$ admits a \sy\ $g$ as in assumption {\sl a)} 
in Theorem 1, and also a \sy\ generated by a linear VF $Lu\cd\grad$ such 
that in addition (31) and (32) are satisfied. Then, if $\^F_1$ has the form 
(30), the NT is \co t. The result can be trivially extended to the 
case that $f$ admits an algebra of (more than one) \sys\ $L_ku\cd\grad$.

\pn
{\bf Example 2.} The same as Example 1, here with $b=0$  and $a_1,a_2\not= 
0$. In the NF now $\mu_1=a_1r^2z+\ldots\not=0$, but $\mu_2$ may be zero. 
If $\mu_2\not= 0$, Theorem 1 can be applied. 
If $\mu_2=0$, then $\^F_1$ has the form (30); 
on the other hand, it is clear that no $SO_2$-invariant \an\ functions are 
simultaneously \coms\ of $\.u=Au$ and $\.u=u$, then all assumptions of 
Theorem 2 are satisfied, and the NT is \co t.

The example given in [9] can be viewed as another example of Theorem 2, 
in the presence of a larger \sy\ (the Lie algebra of the group $SO_3$).
\bs
To conclude, it can be interesting to point out the following peculiar 
property of the present approach. 
\pn
{\bf Remark 3.} All the results in this paper
are peculiar of {\it non-Hamiltonian} DS: indeed, an Hamiltonian DS never 
satisfies the crucial hypothesis, i.e. assumption {\sl b)} of Theorem 1. Let 
in fact $H=H(u)$, with $u\equiv(q,p)\in\R^{2m}$, be an \an\ Hamiltonian and
$$\.u=J\grad H$$
be the associated DS, where $J$ is the symplectic matrix. Writing 
$H=H_0+H_R$, where $H_0$ is the quadratic part of $H$, we have clearly
$Au=J\grad H_0$ and $F(u)=J\grad H_R$, and the requirement that $F(u)$ is 
in NF, $\^F(u)=J\grad\^H_R$, becomes now the requirement that $H_0$ is a 
\com\ of $\^H_R$ (cf. [15,16]). Then, eq. (15) always admits  nontrivial 
solutions of the form
$$S=\eta(H_0)\^F_1$$
for any (regular) function $\eta$ of $H_0$.

\vfill\eject

\baselineskip 0.5 cm 
\Ref

[1] Bruno A.D., 1971 "Analytical form of differential equations", Trans. 
Moscow Math. Soc. {\bf 25}, 131; and  {\bf 26}, 199 

[2] Bruno A.D., 1989, "Local methods in nonlinear differential equations",
Springer, Berlin  

[3] Arnold V.I., 1988, "Geometrical methods in the theory of
differential  equations", Springer, Berlin 

[4] Arnold V. I. and  Il'yashenko Yu. S., 1988 {\it Ordinary differential
equations}; in:  Encyclopaedia of Mathematical Sciences - vol. I,
Dynamical Systems I, (D.V. Anosov and V.I. Arnold, eds.), pp. 1-148;
Springer, Berlin

[5]  Olver P.J., 1986, "Applications of Lie groups to differential
equations", Springer, Berlin

[6] Ovsjannikov L.V., 1962, "Group properties of differential equations",
Novosibirsk; (English transl. by Bluman G.W., 1967); and 1982,
"Group analysis of differential equations", Academic Press, New York

[7] Cicogna G. and  Gaeta G., 1994,  Journ. Phys. A: Math. Gen. 
{\bf 27}, 461 and 7115; and 1997, IHES preprint

[8] Bruno A.D. and Walcher S., 1994, J. Math. Anal. Appl. {\bf 183}, 571

[9] Cicogna G., 1995, Journ. Phys. A: Math. Gen. {\bf 28},  L179,
and 1996, Journ. Math. Anal. Appl. {\bf 199}, 243

[10] Gaeta G. and Marmo G., 1996,  Journ. Phys. A: Math. Gen. {\bf 29}, 5035
 
[11] Bambusi D., Cicogna G., Gaeta G., and Marmo G., 1997,  preprint

[12] Elphick C., Tirapegui E., Brachet M. E., Coullet P., and  
Iooss G., 1987, Physica {\bf D, 29}, 95

[13] Walcher S., 1991, Math. Ann. {\bf 291},  293

[14] Iooss G. and Adelmeyer M., 1992, {\it Topics in bifurcation theory and
applications}; World Scientific, Singapore 

[15] Moser J.K.,  1968, Mem. Amer. Math. Soc. {\bf 81}, 1 

[16] Arnold V. I., Kozlov  V.V., and Neishtadt  A. I., 1993, {\it Mathematical
aspects of classical and celestial mechanics}; in:  Encyclopaedia
of Mathematical Sciences - vol. 3, Dynamical Systems III, (V.I. Arnold
ed.), pp. 1-291; Springer, Berlin 

\bye